# Structure and properties of α- and β- CeCuSn: A single-crystal and Mössbauer spectroscopic investigation

C. Peter Sebastian, Sudhindra Rayaprol, Rolf-Dieter Hoffmann, Ute Ch. Rodewald, Tania Pape, Rainer Pöttgen[*]

*Institut für Anorganische und Analytische Chemie and NRW Graduate School of Chemistry, Universität Münster, Corrensstrasse 30, D-48149 Münster, Germany*

**Abstract**

*Two modifications of CeCuSn were prepared from the elements: the high-temperature (β) modification crystallizes directly from the quenched sample, while the low-temperature (α) modification forms after annealing at 700 °C for one month. Both modifications were investigated by X-ray powder and single crystal diffraction. We find for β-CeCuSn a structure of ZrBeSi type, space group $P6_3/mmc$, a = 458.2(1), c = 793.7(2) pm, wR2 = 0.0727, 148 $F^2$ values, 8 variable parameters. In the case of α-CeCuSn we find the NdPtSb type structure, space group $P6_3mc$, a = 458.4(1), c = 785.8(2) pm, wR2 = 0.0764, 233 $F^2$ values, 11 variable parameters. The copper and tin atoms build up layers of ordered [$Cu_3Sn_3$] hexagons. The layers are planar in β-CeCuSn, however, with highly anisotropic displacements of the copper and tin atoms. In α-CeCuSn a puckering effect is observed resulting in a decrease of the c lattice parameter. Both modifications of CeCuSn exhibit antiferromagnetic ordering, however, there is a considerable difference in their magnetic behaviour. We show the anomalies in the physical properties of the α- and β- modifications of CeCuSn by Mössbauer spectroscopy, magnetic and specific heat measurements and explain their structure-property relations.*

## 1. INTRODUCTION

The *R*CuSn stannides (*R* = rare earth element) have been studied intensively in recent years with respect to their interesting magnetic and electrical properties. An overview of the literature is given in [1]. If the rare earth elements are trivalent, the *R*CuSn stannides crystallize with superstructures [2] that derive from the well known AlB$_2$ type. The hexagonal structures have AB-AB stacking sequences of planar (ZrBeSi type [3]) or puckered (NdPtSb [4] or LiGaGe [5, 6] type, depending on the degree of puckering) [$Cu_3Sn_3$] hexagons, which are rotated by 60° in every other layer. The degree of puckering strongly depends on the size of the rare earth element. With lanthanum almost planar layers have been observed [1, 7], while ScCuSn shows a strong puckering, leading to a slightly elongated tetrahedral [CuSn] network [1].

CeCuSn has most intensively been investigated [8–21]. But the structure of CeCuSn has so far only been investigated on the basis of powder diffraction data, i.e. laboratory X-ray, synchrotron, and neutron powder diffraction. In the earlier work by Dwight [8], Marazza [14], and Riani [16] the CaIn$_2$ type (space group $P6_3/mmc$) with a statistical distribution of copper and tin was assumed, while an ordered arrangement of copper and tin was reported by Yang et al. [10] and Adroja et al. [15]. The best powder data (neutron data) have recently been reported by Weill et al. [21]. They clearly established the ordering and a slight puckering of the [$Cu_3Sn_3$] hexagons. Concerning the sample preparation, two different routes have been established: (i) arc-melting of the elements directly followed by annealing in sealed silica tubes between 750 and 850 °C for up to four weeks or (ii) crystal growth via the Czochralski technique.

Magnetic, resistivity and specific heat studies on single-crystal and polycrystalline CeCuSn exhibits complex magnetic ordering in this compound [11-12, 20]. Sakurai *et al.* observed antiferromagnetic ordering ($T_N$) around 8 K in arc-melted CeCuSn, which also exhibited weak spontaneous magnetization. However the samples (which were eventually polycrystalline) prepared by the Czochralski pulling method exhibited $T_N$ around 10 K without any spontaneous magnetization [11]. Specific heat studies of Yang *et al.* exhibits two magnetic transitions around 8.6 and 7.3 K [10]. These anomalies in CeCuSn were further supported by the µSR studies by Kalvius *et al.*, who observed the presence of two magnetic states below 11 K [17]. A recent single-crystal neutron diffraction experiment shows that the onset of antiferromagnetic ordering occurs around 12 K and an inflection around 8 K is seen in the temperature dependence of the magnetic intensities. The magnetic ground state is characterized by a magnetic wave vector **k** = (0.115, 0, 0) [20]. CeCuSn has repeatedly been studied in order to understand the ground-state properties [19].

The complex magnetic behaviour of CeCuSn motivated us to investigate the magnetic structure of this interesting compound using the $^{119}$Sn Mössbauer spectroscopy. However during the synthesis, owing to the synthetic conditions, we observed two structural modifications of this compound.

The first method of synthesis was arc-melting the elements, and annealing a part of this *as cast* sample to obtain the second sample. Firstly β-CeCuSn was obtained by arc-melting the elements in ideal atomic ratio and quenching the ingot. This sample revealed a significantly larger *c* parameter and magnetic ordering behaviour different than the ones reported in the literature [10-12, 15-17, 20]. The second structural modification, α-CeCuSn was obtained after annealing the first sample (i.e., β-CeCuSn) in a sealed silica tube at 700 °C for one month. The lattice parameters for this sample were close to the values reported in the literature. In order to determine the actual crystal structure of CeCuSn, we report herein a precise single crystal study on the high- (β) and low-temperature (α) modifications of CeCuSn. We probe the relation between the structures and their physical properties by the magnetic, specific heat and $^{119}$Sn Mössbauer spectroscopic studies on polycrystalline samples of both modifications.

## 2. Experimental
*2.1. Syntheses*

Starting materials for the preparation of the CeCuSn samples were a larger cerium ingot (Johnson Matthey), copper wire (Johnson Matthey, ∅ 1 mm), and tin granules (Heraeus), all with stated purities better than 99.9 %. In the first step, pieces of the cerium ingot were arc-melted [22] to small buttons under an argon atmosphere of ca. 800 mbar. The argon was purified before over titanium sponge (900 K), silica gel, and molecular sieves. The pre-melting procedure strongly reduces shattering during the subsequent reaction with copper and tin. The cerium buttons were then mixed with pieces of the copper wire and the tin granules in the ideal 1:1:1 atomic ratio and all were reacted together in the arc-melting furnace under argon (800 mbar). The product button was melted three times to ensure homogeneity. The total weight losses after the arc-melting procedures were all smaller than 0.5 weight-percent. As defined above, we call this as cast CeCuSn, high temperature modification, i.e., β-CeCuSn. β-CeCuSn was obtained in an amount of 1 g. Part of the sample was subsequently sealed in an evacuated silica tube and annealed at 700 °C for four weeks to obtain the low temperature modification i.e., α-CeCuSn. Compact pieces of polycrystalline α- and β-CeCuSn are stable in air over months while small single crystals and the fine grained grey powder deteriorate in humid air. Both CeCuSn samples were preserved in Schlenk tubes prior to the investigations. Single crystals of CeCuSn exhibit metallic lustre.

*2.2. X-ray Powder and Single Crystal Diffraction*

The arc-melted (β-CeCuSn) and annealed (α-CeCuSn) samples were checked through powder patterns taken in a Guinier camera (equipped with an image plate system Fujifilm, BAS–1800) using Cu $K_{\alpha 1}$ radiation and α-quartz ($a$ = 491.30, $c$ = 540.46 pm) as an internal standard. *No impurities were found in both samples up to the level of X-ray detection*. The hexagonal lattice parameters (Table 1) were obtained by least-squares refinements of the powder data. To ensure correct indexing, the observed patterns were compared to calculated ones [23] using the positional parameters obtained from the structure refinements. The lattice parameters of α-CeCuSn obtained in the present study are in good agreement with the various data reported in literature [8–21]. The powder patterns of β-CeCuSn principally showed broader reflections (exhibiting lower crystallinity) as compared to those of α-CeCuSn. However, in addition, a small anisotropy concerning reflections with odd $l$ indices was observed; i.e. *hkl* reflections were broader than *hk0* reflections in a close 2θ range.

Small, irregularly shaped single crystals were isolated from the quenched and the annealed samples by mechanical fragmentation. The crystals were first examined with white Mo radiation on a Buerger precession camera (equipped with an image plate system, Fujifilm, BAS–1800) in order to establish suitability for intensity data collection. Intensity data of a crystal from the quenched arc-melted sample were collected at room temperature by use of a four-circle diffractometer (CAD4) with graphite monochromatized Mo$K_\alpha$ radiation (71.073 pm) and a scintillation counter with pulse height discrimination. The scans were performed in the *ω/2θ* mode. An empirical absorption correction was applied on the basis of Ψ-scan data followed by a spherical absorption correction.

Additionally, low-temperature intensity data of the crystal from the quenched arc-melted sample were collected at –150°C on a Bruker AXS Smart Apex I with a rotating anode (graphite monochromatized Mo$K_\alpha$; 50 kV, 140 mA) and a fixed detector distance of 50 mm. Ω-scans were employed with a step width of 0.3° at consequent three φ-positions (0, 120, and 240°), thus allowing a semi-empirical absorption correction. The Ω-range was 180° i.e. –90° ≤ θ ≤ 90°. A 30 s counting time was used for each frame. The data acquisition and the absorption correction were carried out with SAINT PLUS 6.28A and SADABS, respectively, supplied by Bruker-AXS. Intensity data of the annealed CeCuSn crystal were collected on a Stoe IPDS-II image plate diffractometer using monochromatized Mo$K_\alpha$ radiation (71.073 pm). A numerical absorption correction was applied to the data. All relevant crystallographic data and details for the three data collections and evaluations are listed in Table 1.

*2.3. Physical Property Measurements and $^{119}$Sn Mössbauer Spectroscopy*

Samples in bulk form were used for the magnetic and specific heat measurements. The magnetic measurements were carried out on QD-PPMS using the VSM insert. The heat capacity ($C_p$) measurements were also performed on the same PPMS (HC option) using *Apiezon N* grease as glue for sticking the sample onto the platform of the puck.

A Ca$^{119m}$SnO$_3$ source was available for the $^{119}$Sn Mössbauer spectroscopic investigations. The samples were placed within a thin-walled PVC container at a thickness of about 10 mg Sn/cm$^2$. A palladium foil of 0.05 mm thickness was used to reduce the Sn K X-rays concurrently emitted by this source. The measurements were conducted in the usual transmission geometry at 78 and 4.2 K.

**3. Results and Discussion**
*3.1. Structure Refinements*

The data collections clearly revealed doubling of the pronounced AlB$_2$ related subcells, as was already evident from the Guinier powder data. Analysis of the systematic extinctions revealed the 6$_3$ screw axis for the three data sets, leading to the possible space groups *P*6$_3$*mc* (NdPtSb type) or *P*6$_3$/*mmc* (ZrBeSi type). The structure of the crystal from the arc-melted and quenched sample was refined first. In view of the recently refined LaCuSn structure [1], the ZrBeSi model, space group *P*6$_3$/*mmc* was refined first. The LaCuSn atomic positions were taken as starting values and the structure was refined using SHELXL–97 (full-matrix least-squares on $F^2$) [24] with anisotropic atomic displacement parameters for all atoms. While the cerium atoms show an almost isotropic displacement, especially the copper atoms behave anisotropically with an extreme displacement in the *c* direction (high $U_{33}$ parameter). This was a clear hint for a violation of the mirror planes at $z = 1/4$ and $z = 3/4$, thus indicating a puckering of the Cu$_3$Sn$_3$ network. In the next step the structure was refined in the non-centrosymmetric space group *P*6$_3$*mc* with puckered [Cu$_3$Sn$_3$] layers. This along with twinning by inversion did not solve the strong anisotropic displacement of the copper atoms. Both the copper and tin atoms remained on the ideal positions at $z = 1/4$ and $z = 3/4$ within two standard uncertainties and the large $U_{33}$ values were as high as in the centrosymmetric refinement. This clearly manifests that in the high-temperature phase microdomains are present which have ordered, slightly puckered Cu$_3$Sn$_3$ hexagons within a single domain, but are out of phase among one another. Anti phase boundaries (APB) are possible in view of the symmetry reduction (*klassengleiche* transition, k2) in going from the AlB$_2$ subcell to the superstructure with ZrBeSi type [2]. Due to the small domain size i.e. high density of APBs along *c* the average structure is described best as ZrBeSi type even though the layers are slightly puckered locally. For the final runs the *P*6$_3$/*mmc* model was assumed again for this crystal.

The same crystal from the arc-melted and quenched sample was then investigated at –150 °C in order to check whether a temperature-induced dynamic can be ruled out. The refinement in space group *P*6$_3$/*mmc* revealed almost the same results, however, with a slightly higher anisotropy for the copper and tin atoms, indicating a stronger tendency for puckering. A refinement with split positions (1/3, 2/3, *z* instead of 1/3, 2/3, 1/4) for the copper atoms did not improve the residuals and further manifests the disorder (short range order) in the quenched sample. We have therefore described this crystal also in space group *P*6$_3$/*mmc*.

Finally the structure of the annealed sample was refined. From the Guinier powder pattern it was already evident, that the *c* lattice parameter of the annealed sample was significantly smaller (793.7 vs. 785.8 pm). This was a clear indication for a puckering and ordering of the [Cu$_3$Sn$_3$] networks. Consequently we have refined the α-CeCuSn structure in space group $P6_3mc$, and the atomic parameters of YCuSn [1] were taken as starting values. The structure refinement went smoothly to the residuals listed in Table 1 and the pronounced anisotropy disappeared. The occupancy parameters were refined in a separate series of least-squares cycles. All sites were fully occupied within two standard uncertainties. A final difference electron-density synthesis was flat and did not reveal any significant residual peaks. The results of the three structure refinement are summarized in Table 1. The atomic coordinates and the interatomic distances are listed in Tables 2 and 3. Further information on the structure refinements is available [25].

*3.2. Crystal Chemistry*

X-ray powder diffraction data revealed significant differences for the *c* lattice parameters, i.e. 793.7(2) pm for the quenched (β-CeCuSn) and 785.8(2) pm for the annealed (α-CeCuSn) sample along with larger FWHM (full width at half maximum) for odd *l* indices. To the best of our knowledge, the short range order high-temperature modification is described here for the first time. All investigations given in literature [8–21] either worked with annealed samples or Czochralski grown single crystals. Most *c* lattice parameters listed in references [8–21] are close to 786 pm, comparable with α-CeCuSn. In some older literature [8, 9, 14, 16], the CaIn$_2$ type with a statistical Cu/Sn occupancy had been assigned to CeCuSn, which is definitely wrong, since in β-CeCuSn ordered [Cu$_3$Sn$_3$] hexagons already occur (Fig 1).

Similar to the recently reported data on LaCuSn [1], CeCuSn shows short range order when quenching the sample from the melt. Due to the large size of the cerium atoms, the [CuSn] networks are well separated from each other and there are no interlayer interactions. After the quenching process the [Cu$_3$Sn$_3$] hexagons are already ordered, however, there is no long-range ordering with respect to the puckering along *c*. We find statistical dislocations above and below the subcell mirror planes, leading to the strong anisotropic displacements. The disorder of the [CuSn] layers among one another occupies space. Consequently we observe a higher *c* lattice parameter and thus a higher cell volume for the high-temperature phase (0.1430 vs. 0.1443 nm$^3$). Therefore APBs are assumed to occur along *c*.

In Figure 2 we illustrate the effect of the apparent disorder suggested by X-ray diffraction. In the high-temperature modification we detect a superposition of both conformational forms present in the low-temperature phase. These are statistically frozen in when quenching the sample from the melt. The microdomains are not resolved by X-ray diffraction and as a consequence we observe the large anisotropic displacement of the copper atoms. In various literature reports on CeCuSn [8–21] the course of the *c* lattice parameter varies from 784.85 [21] to 793.7 pm (this work for β-CeCuSn). Based on the present results we assume that all samples with the higher *c* values have some disorder between the adjacent [CuSn] layers; i.e. varying densities of APBs.

In contrast, the annealed sample clearly shows the ordered NdPtSb type structure, similar to those *R*CuSn stannides with the heavier rare earth elements [1, and ref. therein]. The space group symmetry is reduced from $P6_3/mmc$ (ZrBeSi type) to $P6_3mc$ (NdPtSb type) via a *translationengleiche* symmetry reduction of index 2 (t2) with a loss of the centre of symmetry [2]. Consequently we observed twinning by inversion for the investigated crystal where the twin domains are large enough for independent coherent scattering. For the α-CeCuSn structure we only observe a small anisotropy of the copper displacements, similar to ScCuSn, YCuSn, and LuCuSn [1], and there are no significant residual peaks in the final difference Fourier synthesis. Our single crystal data are in excellent agreement with a recent neutron powder diffraction study [21], but with better precision for the positional parameters. For a more detailed discussion on the chemical bonding and the dimensionality of the [CuSn] networks within the series of *RE*CuSn stannides we refer to the previous work [1, 2].

*3.3 Physical Property Measurements*

In Figure 3 we show the susceptibility ($\chi$ = M/H) plotted as $\chi^{-1}$ vs. T for both modifications of CeCuSn, measured in an applied field of 10 kOe. The curvature in the plots of $\chi^{-1}$ vs. T has been observed in previous studies also [10-11 and 13]. The susceptibility could be fitted above 100 K, to a Curie-Weiss law modified by adding a temperature-independent term and given as $\chi = \chi_0 + C/(T - \theta_p)$, where $\chi_0$ is the temperature-independent Pauli contribution, C is the Curie constant and $\theta_p$ is the paramagnetic Curie temperature. The effective Bohr magnetons number ($\mu_{eff}$) can thus be calculated from the Curie constant [26]. The values of $\chi_0$, $\theta_p$ and $\mu_{eff}$ for α-CeCuSn are 4.15 x $10^{-3}$ emu/mol, 10 K and 2.49 $\mu_B$/Ce atom respectively. Similarly for β-CeCuSn these values are 2.2 x $10^{-3}$ emu/mol, 19 K and 2.46 $\mu_B$/Ce atom. The values of $\mu_{eff}$ for both compounds are in close agreement with the value of the effective Bohr magnetons number for the free $Ce^{3+}$, 2.54 $\mu_B$, indicating trivalent cerium in the α- and β- modification of CeCuSn. The positive values of $\theta_p$ indicate ferromagnetic interactions.

A very clear difference in the magnetism of these two compounds can be seen from the low field susceptibility curves plotted in Fig. 4. The top panel shows the zero field cooled (ZFC) and field cooled (FC) $\chi$(T) curves for α-CeCuSn measured in a field of 100 Oe. The earlier heat capacity studies of annealed CeCuSn show two magnetic transitions at 8.6 and 7.4 K [12]. Consistent with these temperatures, there is a sudden rise in $\chi$(T) of α-CeCuSn around 8.6 K as if it is an ferromagnetic ordering *vide infra*. The observation of a ferromagnetic component is consistent with the observations made by Sakurai *et al.* for CeCuSn (high value of residual resistivity, mixing of small ferromagnetic component). The positive value of $\theta_p$ however contradicts the observation of negative $\theta_p$ by Yang *et al.* [10, 11]. The ZFC-FC bifurcates around 7.2 K, and the ZFC undergoes antiferromagnetic transition with a broad peak around 6.3 K. In the bottom panel of Fig. 4, the ZFC-FC curves for β-CeCuSn are shown. The features are totally different from the one seen for α-CeCuSn. The susceptibility in ZFC and FC state increases with decreasing temperature and exhibits a broad peak below 10 K, which can be attributed as short-range magnetic ordering [27]. This feature is directly correlated to the missing long-range structural ordering of the $Cu_3Sn_3$ hexagons in β-CeCuSn (see Figure 2).

The difference in the magnetism of α-CeCuSn and β-CeCuSn can also be seen in the behaviour of the M(H) curves. In Figure 5 we have shown the magnetization as a function of field up to 80 kOe at 5 K. It is interesting to observe the occurrence of small steps in the magnetization of α-CeCuSn at fields of 8, 10, 20 and 30 kOe both in up (field ramping up) and down (field ramping down) cycles of field ramping, indicating spin reorientation effects. A small remanent magnetization is seen at low fields and can be more clearly seen in the hysteresis loop (discussed below). M(H) at 5 K for β-CeCuSn is shown in the bottom panel of Fig. 5. M increases with H without any steps in the magnetization as was observed in the case of α-CeCuSn. The M(H) curve clearly resembles features of an antiferromagnet, i.e., non-linear increase in M with increasing H with a tendency to saturate at higher fields. The saturation moment value at 80 kOe, for both compounds is about 1 $\mu_B$ per mole which is usually observed for ternary trivalent cerium compounds with localized *4f* magnetism [28]. The reduced moment values, from the expected $g_J \times J = 2.14$ $\mu_B$/mol for $Ce^{3+}$ can be attributed to the crystal field splitting effects on the $J = 5/2$ ground state of the $Ce^{3+}$ ion. Some of the compounds exhibiting saturation moment values (in units of $\mu_B$) similar to CeCuSn are, 1.09 - CeAuGe [29], 1.2 – CePdSb [30], or 1.12 – CeAgGa [31].

A hysteresis loop measured at 5 K for α-CeCuSn up to ±10 kOe is shown as the inset in the top panel of Fig. 5. A small remanent field, few Oesterds wide can be clearly seen. The loop collapses when the field crosses this critical value. Such a hysteresis loop is usually seen for compounds exhibiting cluster-glass like behaviour [32]. This means that there is a presence of small ferromagnetic component in this compound. Therefore it can be assumed that in α-CeCuSn there exists along with antiferromagnetic ordering, a spontaneous magnetisation at low temperatures resulting in formation of cluster-glass like anomalies. This phenomenon for α-CeCuSn is not clearly understood at present and needs further detailed experimental investigations. For the β-CeCuSn, shown in the inset of the bottom panel of Fig. 5, there is no hysteresis at 5 K, and the curve is almost linear with almost no broadening around the origin. In the context of different magnetic behaviour it is very interesting to see that introducing long range order of the puckering in the [CuSn] network significantly alters the magnetism of CeCuSn.

Now we focus on the specific heat behaviour of the α- and β- forms of CeCuSn in Fig. 6. The peaks in C(T) of α-CeCuSn are observed at 8.6(1) and 7.5(1) K which are consistent with the observation of two transitions in the $C_p$ data of the annealed CeCuSn samples [10, 12]. We consider 8.6 K as $T_N$ in accordance with the previous $C_p$ studies on CeCuSn [12]. There is a small anomaly around 10 K, consistent with the earlier reports that the onset of antiferromagnetic ordering takes place around this temperature. The C(T) for β-CeCuSn is shown in the bottom panel of Fig. 6. Contrary to α-CeCuSn, there is only one step like anomaly around 7.9(1) K which we consider as $T_N$ for β-CeCuSn. There are no other visible anomalies in the C(T) curve of β-CeCuSn. It may be recalled that the magnetic phase diagram of CeCuSn given by Nakotte *et al* indicates that magnetic anisotropy is intrinsic to the CeCuSn compound [12].

In order to understand the ground state of CeCuSn, it is important to get the magnetic part of the heat capacity for both modifications. Structurally, YCuSn [1] is similar to α-CeCuSn and LaCuSn [1] is close to β-CeCuSn, therefore we have taken them as non-magnetic reference for subtracting the lattice part and to obtain the magnetic part of heat capacity for α- and β-CeCuSn. The C(T) curves for YCuSn and LaCuSn are also shown in the top and bottom panels, respectively. The magnetic part of the heat capacity for α- and β-CeCuSn is shown in different forms in Fig. 7. A λ-anomaly corresponding to the magnetic transitions are clearly seen in both the forms of CeCuSn. The difference in magnetic behaviour of the two forms of CeCuSn is clearly evident. In α-CeCuSn the two peaks corresponding to the ordering temperatures are clearly seen in $C_{mag}$ with a sudden rise around 12 K. For β-CeCuSn small peaks around 8 and 13 K are seen in $C_{mag}$. In both compounds a small bump in $C_{mag}$ is seen at 24(1) K. A careful look at Fig. 5 of Ref. 10 shows that the curves of $C_p/T$ vs. T measured in different applied fields crosses each other around 25 K, indicating heavy fermion like behaviour in CeCuSn [10, 33].

The plot of $C_{mag}/T$ vs. $T^2$ for the entire temperature range of measurement is shown in the main panel of Fig. 8. To highlight the low temperature part, i.e., to clearly see the $T^3$ behaviour of $C_{mag}$, we have plotted $C_{mag}/T$ vs. $T^2$ on an expanded scale as an inset. In the ordered region, below 5 K, the $C_{mag}$ follows $T^3$ and from the fit of this linear region we obtain the electronic coefficient of heat capacity ($\gamma$) to be 140 and 132 mJ/mol K$^2$ for α-CeCuSn and β-CeCuSn, respectively. The value of $\gamma >$ 100 mJ/mol K$^2$ puts both forms of CeCuSn in the class of moderate heavy fermion materials.

*3.4. $^{119}$Sn Mössbauer Spectroscopy*

$^{119}$Sn Mössbauer spectra of the α- and β-CeCuSn samples measured at 78 and 4.2 K, are shown in Figures 9 and 10, respectively. In agreement with the crystal structures the spectra are easily understood in terms of single crystallographic tin sites. The hyperfine parameters derived at the above mentioned temperatures from the recorded spectra are presented in Table 4. As expected, above the $T_N$, the non-cubic tin site symmetry (3$m$.) in α-CeCuSn results in a small quadrupolar splitting 0.45(3) mm/s which agrees well with the values measured for the Sn site in the isostructural compounds $R$CuSn ($R$ = Gd-Er) [34]. In contrast, the quadrupole splitting value of 0.32(8) mm/s measured for β-CeCuSn is significantly smaller, consistent with the higher Sn site symmetry ($\bar{6}m2$) in this compound which is close to the value of LaCuSn [1]. The smaller quadrupole splitting for α-CeCuSn indicates a decrease in the electric field gradient at the tin site in α-CeCuSn as compared to β-CeCuSn. The isomer shifts are in the typical range observed for tin in intermetallic compounds [35-39] and comparable to those measured for the corresponding gold and silver stannides.

The Mössbauer spectra at 4.2 K (Figure 10) show magnetic hyperfine field splitting below $T_N$ for α- and β-CeCuSn. The transferred hyperfine fields are 1.03(1) and 1.97(4) T for α- and β-CeCuSn, respectively. Since the values for the transferred hyperfine fields are small, and thus correlated with the line width parameter, the absolute values should not be overinterpreted. They suggest a structurally frustrated cerium in β-CeCuSn.

**4. Summary**

We were able to obtain two structural modifications of CeCuSn by varying the synthesis conditions (β modification was obtained by quenching the arc-melted sample and α modification by annealing the quenched sample). The successful analysis of single crystal data gives precise information about the structure of CeCuSn, which will help in interpreting the anomalies observed in physical properties. It is therefore clear in the case of CeCuSn that by quenching the sample, we obtain a short range order of locally slightly puckered [CuSn] layers. Anti phase boundaries (APB) disturb the order of the layers in the c-direction. Annealing reduces the effect of APBs completely, the puckering becomes stronger and three-dimensional long range order is present which strongly affects the magnetism of this compound. The magnetic and specific heat studies on α-CeCuSn exhibits complex magnetic ordering. This compound exhibits closely spaced steps in magnetization measurement at 5 K and exhibits signatures of spin-glass with small remanent field around origin in hysteresis loop. On the other hand, β-CeCuSn behaves as a antiferromagnet with no steps in the M(H) curve. The differences in magnetic behaviour of α- and β-CeCuSn are therefore believed to be the manifestation of their structural differences.

We believe that the results presented here will simulate further work to completely understand the 'complex' magnetic behaviour of CeCuSn.

*Acknowledgments*
This work was financially supported by the Deutsche Forschungsgemeinschaft. C.P.S. and S.R. are indebted to the NRW Graduate School of Chemistry and to the Alexander von Humboldt-Foundation respectively for research grants.

Table 1    Crystal data and structure refinement for CeCuSn , molar mass = 322.35g/mol, Z = 2, F(000) = 274

| Preparation | quenched | | annealed |
|---|---|---|---|
| Temperature | 22° C | −150° C | 22° C |
| Instrument | CAD4 | Smart Apex I | IPDS II |
| Unit cell dimensions | $a$ = 458.2(1) pm | $a$ = 455.8(2) pm | $a$ = 458.4(1) pm |
| (Guinier powder data) | $c$ = 793.7(2) pm | $c$ = 791.6(6) pm | $c$ = 785.8(2) pm |
|  | $V$ = 0.1443 nm$^3$ | $V$ = 0.1424 nm$^3$ | V = 0.1430 nm$^3$ |
| Space group | $P6_3/mmc$ | $P6_3/mmc$ | $P6_3mc$ |
| Calculated density | 7.42 g/cm$^3$ | 7.52 g/cm$^3$ | 7.49 g/cm$^3$ |
| Crystal size | 25 x 45 x 65 μm$^3$ | | 20 x 20 x 70 μm$^3$ |
| Transm. ratio (max/min) | 1.40 | 1.36 | 1.77 |
| Absorption coefficient | 31.0 mm$^{-1}$ | 31.4 mm$^{-1}$ | 31.3 mm$^{-1}$ |
| θ range | 5° to 35° | 5° to 31° | 5° to 34° |
| Range in $hkl$ | ±7, ±7, ±12 | ±6, −6/+3, -8/+11 | ±7, ±7, ±11 |
| Total no. reflections | 2334 | 963 | 1895 |
| Independent reflections | 148 ($R_{int}$ = 0.0569) | 102 ($R_{int}$ = 0.0526) | 233 ($R_{int}$ = 0.0301) |
| Reflections with I>2σ(I) | 129($R_{sigma}$ = 0.0161) | 88 ($R_{sigma}$ = 0.0277) | 207 ($R_{sigma}$ = 0.0201) |
| Data/parameters | 148 / 8 | 102 / 8 | 233 / 11 |
| Goodness-of-fit on F$^2$ | 1.200 | 1.166 | 1.106 |
| Final R indices [I>2σ(I)] | R1 = 0.0332 | R1 = 0.0376 | R1 = 0.0282 |
|  | $w$R2 = 0.0705 | $w$R2 = 0.0792 | $w$R2 = 0.0752 |
| R indices (all data) | R1 = 0.0397 | R1 = 0.0499 | R1 = 0.0354 |
|  | $w$R2 = 0.0727 | $w$R2 = 0.0830 | $w$R2 = 0.0764 |
| Extinction coefficient | 0.012(2) | 0.0002(18) | 0.013(3) |
| BASF | – | – | 0.55(10) |
| Largest diff. peak and hole | 6.08 and –2.02 e/Å$^3$ | 4.56 and –2.98 e/Å$^3$ | 2.00 and –2.16 e/Å$^3$ |

Table 2  Atomic coordinates and anisotropic displacement parameters ($pm^2$) for CeCuSn. $U_{eq}$ is defined as one third of the trace of the orthogonalized $U_{ij}$ tensor. $U_{11} = U_{22}$, $U_{13} = U_{23} = 0$.

| Atom | Wyckoff position | x | y | z | $U_{11}$ | $U_{33}$ | $U_{12}$ | $U_{eq}$ |
|---|---|---|---|---|---|---|---|---|
| *Quenched, $22^0C$, $P6_3/mmc$* | | | | | | | | |
| Ce | 2a | 0 | 0 | 0 | 127(4) | 93(5) | 64(2) | 116(3) |
| Cu | 2c | 1/3 | 2/3 | 1/4 | 108(8) | 789(32) | 54(4) | 335(10) |
| Sn | 2d | 1/3 | 2/3 | 3/4 | 69(4) | 172(6) | 35(2) | 103(4) |
| *Quenched,$-150^0C$, $P6_3/mmc$* | | | | | | | | |
| Ce | 2a | 0 | 0 | 0 | 83(6) | 90(8) | 42(3) | 86(6) |
| Cu | 2c | 1/3 | 2/3 | 1/4 | 51(12) | 1145(56) | 26(6) | 416(18) |
| Sn | 2d | 1/3 | 2/3 | 3/4 | 26(7) | 170(10) | 13(3) | 74(6) |
| *Quenched ,annealed at $700^0C$, $22^0C$, $P6_3mc$* | | | | | | | | |
| Ce | 2a | 0 | 0 | 0[a] | 113(3) | 95(5) | 57(1) | 107(3) |
| Cu | 2b | 1/3 | 2/3 | 0.2733(7) | 105(5) | 401(45) | 53(3) | 204(15) |
| Sn | 2b | 1/3 | 2/3 | 0.7343(5) | 72(3) | 192(10) | 36(1) | 112(4) |

[a]The cerium atom was kept fixed at the origin of the cell.

Table 3  Interatomic distances (pm), calculated with the lattice parameters taken from X-ray powder data of CeCuSn. All distances within the first coordination spheres are listed. Standard deviations are equal or less than 0.1 pm.

| Quenched, 22°C | | | | Quenched, -150°C | | | | Quenched, annealed, 22°C | | | |
|---|---|---|---|---|---|---|---|---|---|---|---|
| Ce: | 6 | Sn | 330.7 | Ce: | 6 | Sn | 329.3 | Ce: | 3 | Cu | 319.0 |
|  | 6 | Cu | 330.7 |  | 6 | Cu | 329.3 |  | 3 | Sn | 322.4 |
|  |  |  |  |  |  |  |  |  | 3 | Sn | 337.1 |
|  |  |  |  |  |  |  |  |  | 3 | Cu | 340.8 |
| Cu: | 3 | Sn | 264.5 | Cu: | 3 | Sn | 263.2 | Cu: | 3 | Sn | 266.4 |
|  | 6 | Ce | 330.7 |  | 6 | Ce | 329.3 |  | 3 | Ce | 319.0 |
|  |  |  |  |  |  |  |  |  | 3 | Ce | 340.8 |
| Sn: | 3 | Cu | 264.5 | Sn: | 3 | Cu | 263.2 | Sn: | 3 | Cu | 266.4 |
|  | 6 | Ce | 330.7 |  | 6 | Ce | 329.3 |  | 3 | Ce | 322.4 |
|  |  |  |  |  |  |  |  |  | 3 | Ce | 337.1 |

Table 4  Fitting parameters of $^{119}$Sn Mössbauer measurement of α- and β-CeCuSn at 78 and 4.2 K. Numbers in parentheses represent the statistical errors in the last digit. (δ), isomeric shift; $\Delta E_Q$, electric quadrupole splitting; $|B_{hf}|$, magnetic hyperfine field; (Γ), experimental line width.

| Compound | Temperature (K) | δ (mm/s) | $\Delta E_Q$ (mm/s) | $B_{hf}$ (T) | Γ (mm/s) |
|---|---|---|---|---|---|
| α-CeCuSn | 78 | 1.94(1) | 0.45(3) | - | 1.12(3) |
|  | 4.2 | 1.99(3) | 0.05(2) | 1.03(1) | 1.98(1) |
| β-CeCuSn | 78 | 1.89(2) | 0.32(8) | - | 1.02(7) |
|  | 4.2 | 1.93(1) | 0.12(2) | 1.97(4) | 1.83(3) |

FIGURE CAPTIONS

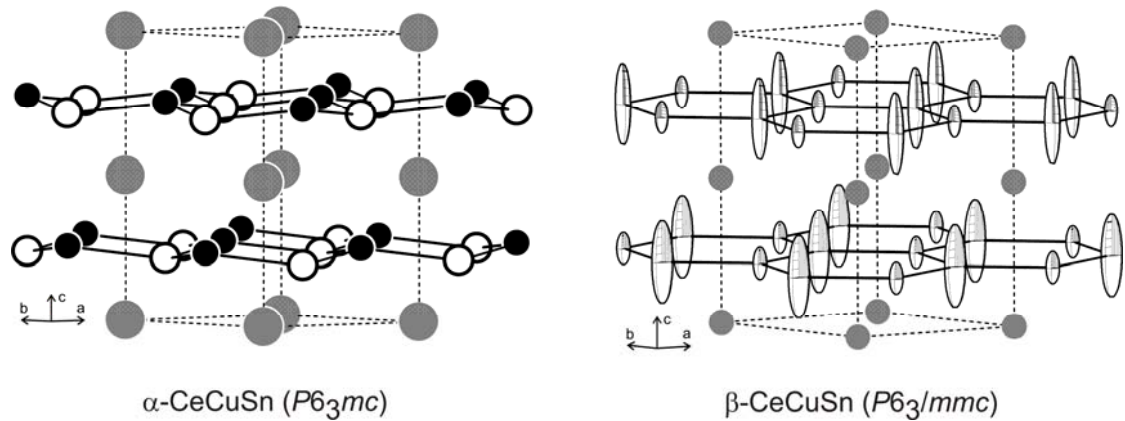

Fig. 1. The crystal structures of α- and β-CeCuSn. The cerium, copper, and tin atoms are drawn as medium grey, filled, and open circles, respectively. The displacement ellipsoids for β-CeCuSn are drawn at the 99 % probability level. The two-dimensional [CuSn] networks are emphasized. For details see text.

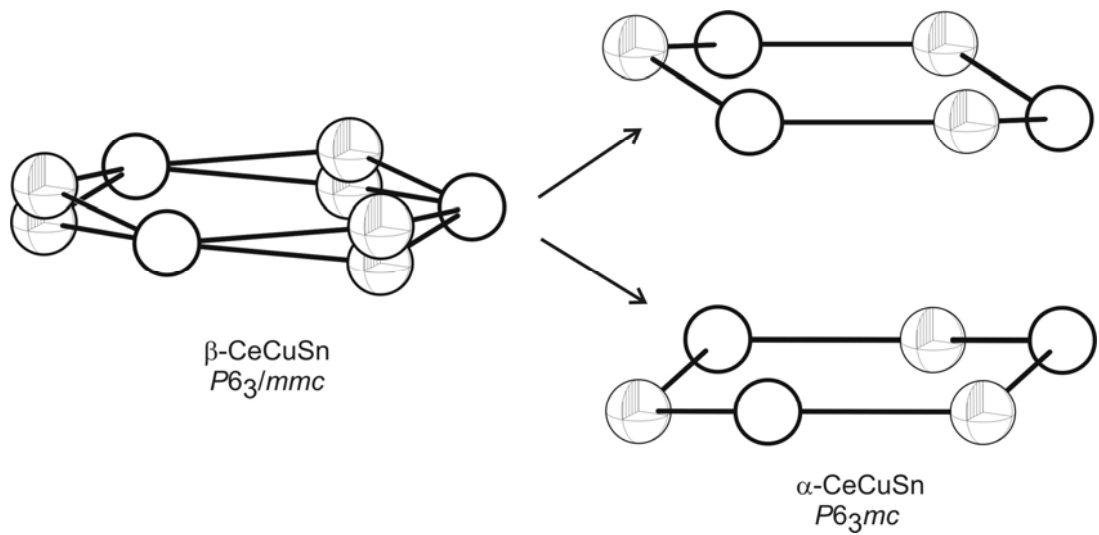

Fig. 2. Cutouts of the α- and β-CeCuSn structures. Only one $Cu_3Sn_3$ hexagon is emphasized. The copper and tin positions are drawn as octants and open circles, respectively. In β-CeCuSn only a superposition of the ordered arrangements of α-CeCuSn is seen by X-ray diffraction. The local picture of α-CeCuSn is valid also for β-CeCuSn, except for a lesser puckering and the occurrence of microdomains. For details see text.

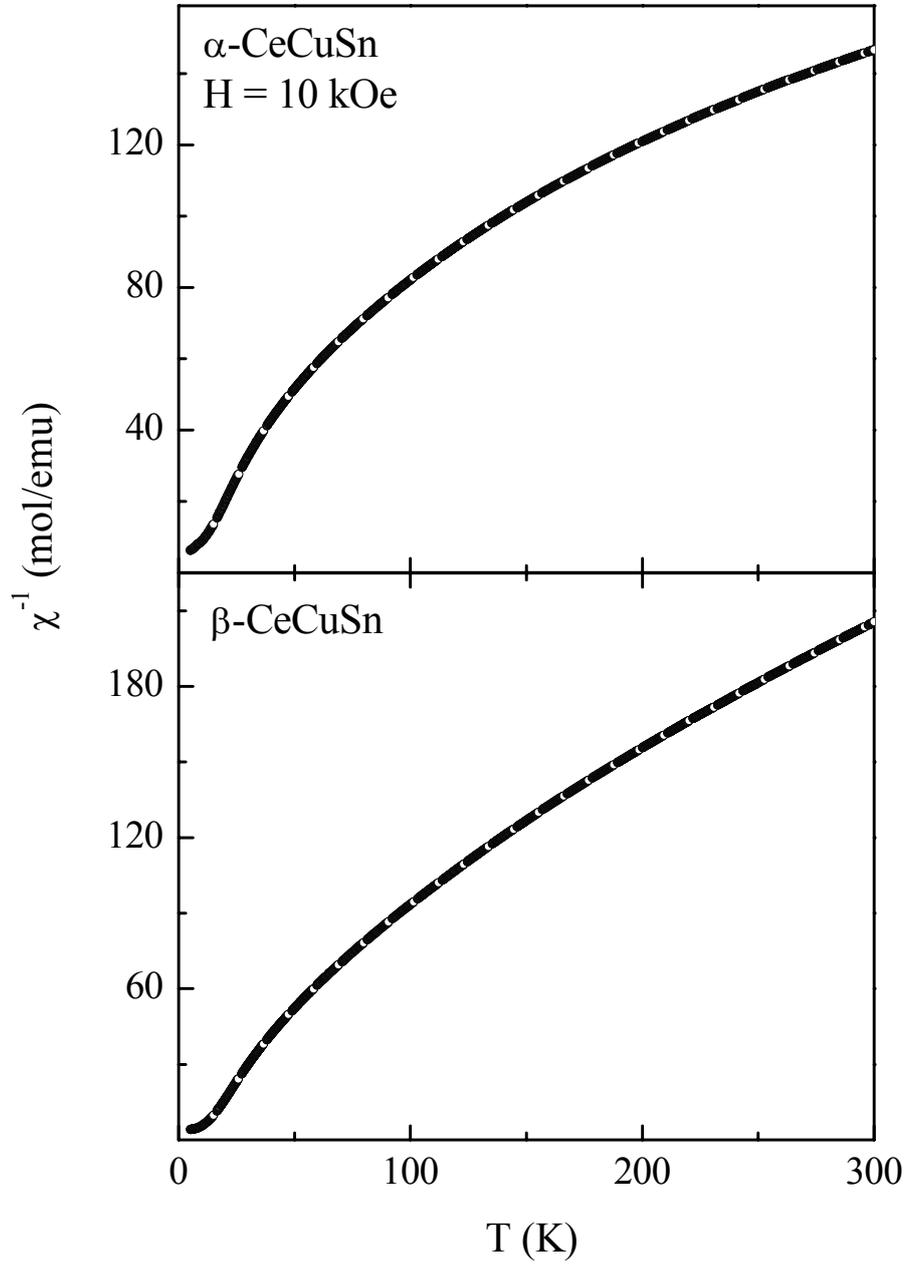

Fig. 3. The inverse susceptibility ($\chi^{-1}$) as a function of temperature for α-CeCuSn (top) and β-CeCuSn (bottom) measured in a field of 10 kOe. The above data was fitted to a modified Curie-Weiss law including a temperature independent term, $\chi_0$. For details see text.

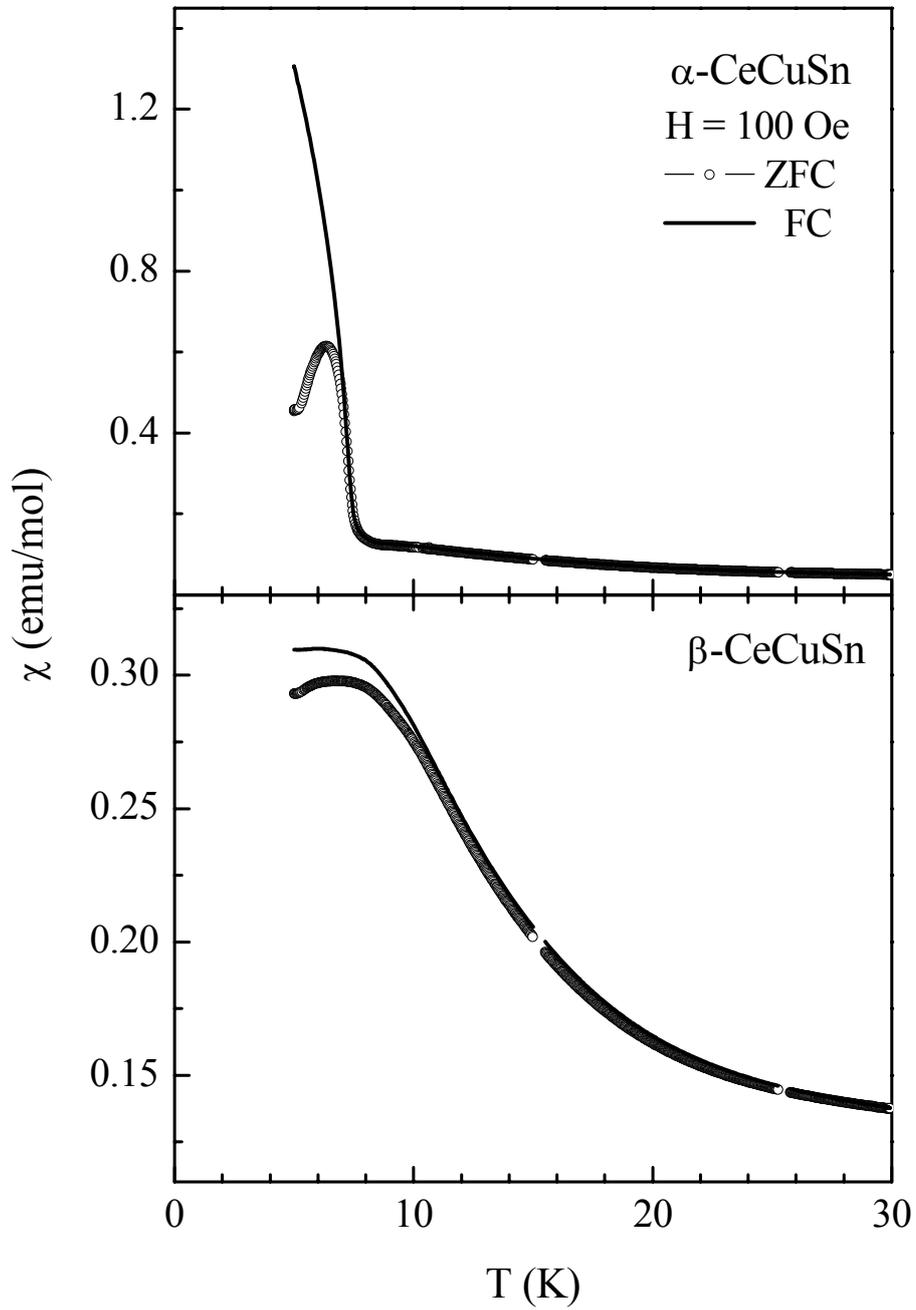

Fig. 4. The susceptibility ($\chi$ = M/H) for α- and β-CeCuSn measured at H = 100 Oe after zero field cooling (ZFC) and field cooling (FC) the sample.

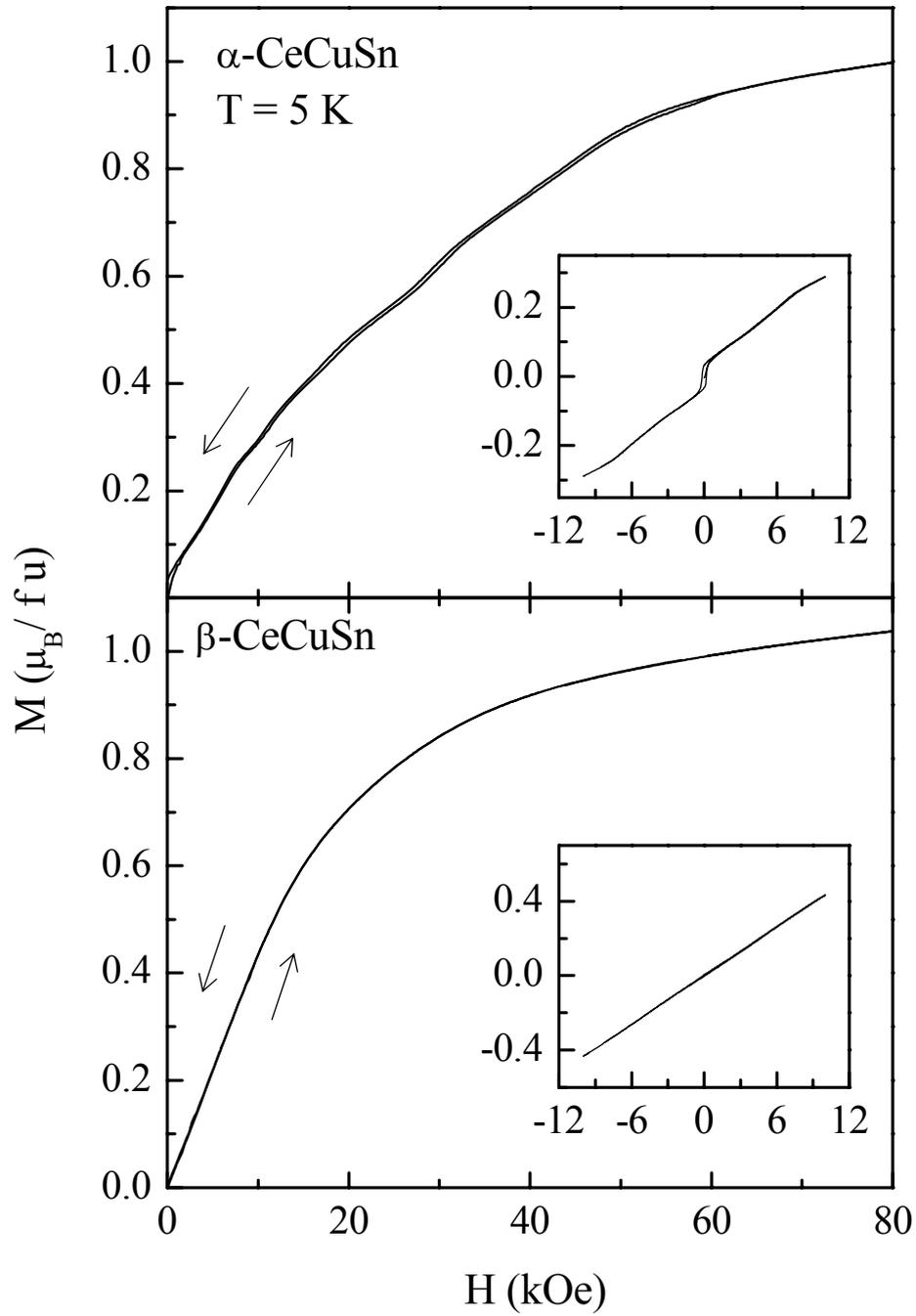

Fig. 5. Magnetization as a function of varying field for α- and β-CeCuSn measured at 5 K. The arrows show the direction of field. The inset in both panels shows the hysteresis loop recorded at the same temperature for the respective compound.

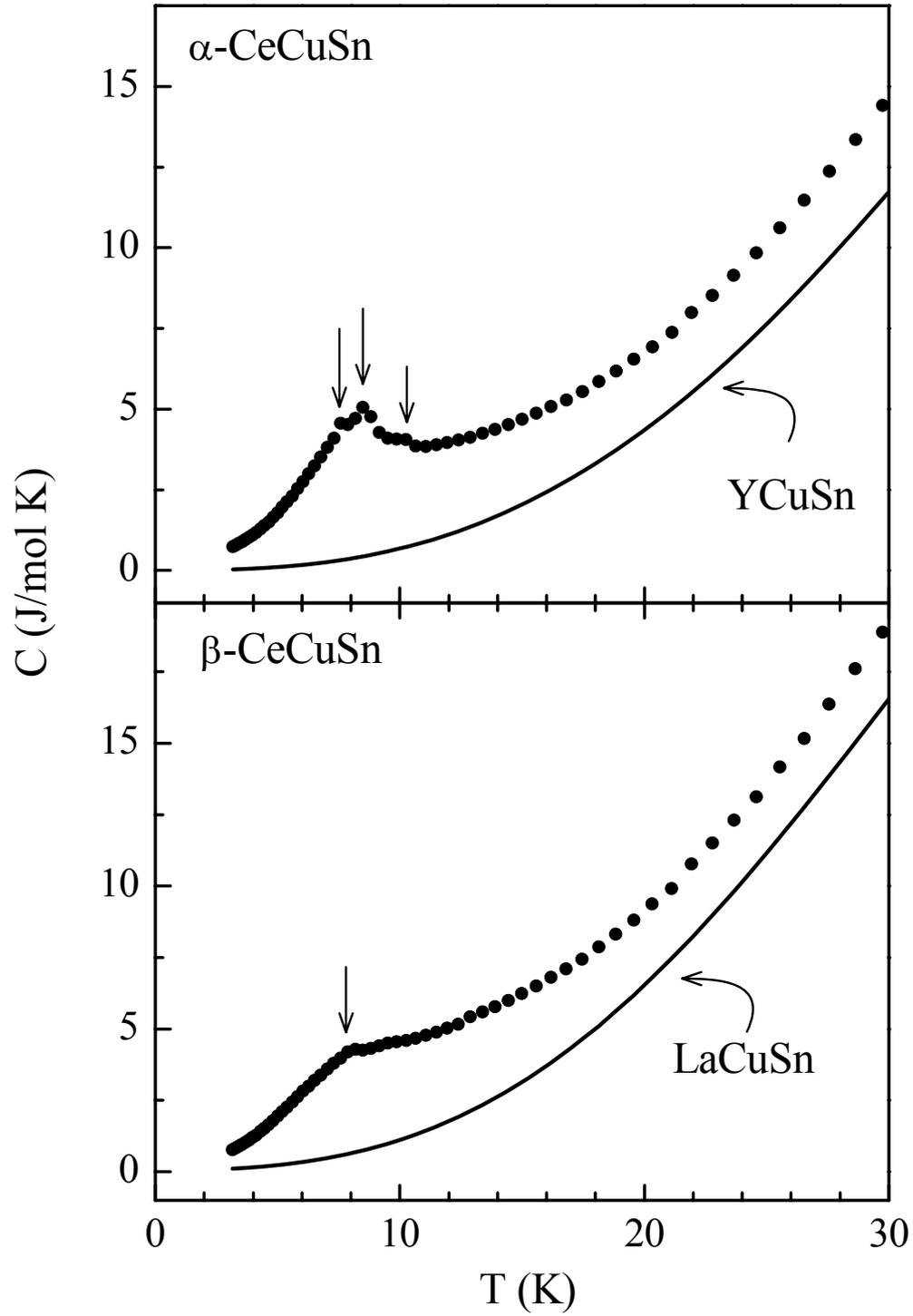

Fig. 6. Total heat capacity (C) for α- and β-CeCuSn. In both panels plotted is the heat capacity of their respective non-magnetic counterparts, YCuSn and LaCuSn.

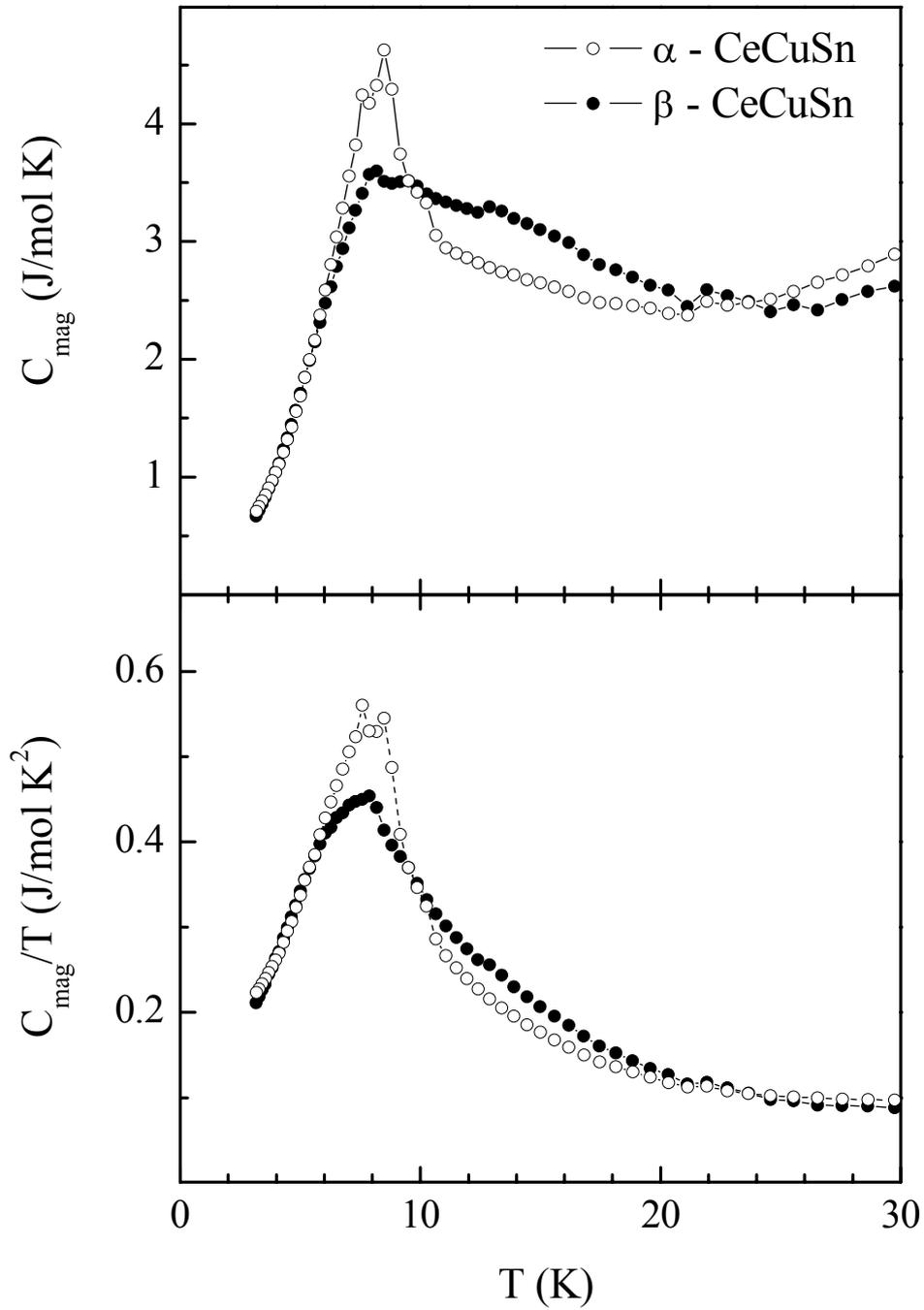

Fig. 7. The magnetic part of the heat capacity ($C_{mag}$) for α- and β-CeCuSn plotted in different ways. $C_{mag}$ is obtained by subtracting the lattice part (i.e., heat capacity of non-magnetic YCuSn and LaCuSn) from the total heat capacity of α- and β-CeCuSn. See text for details.

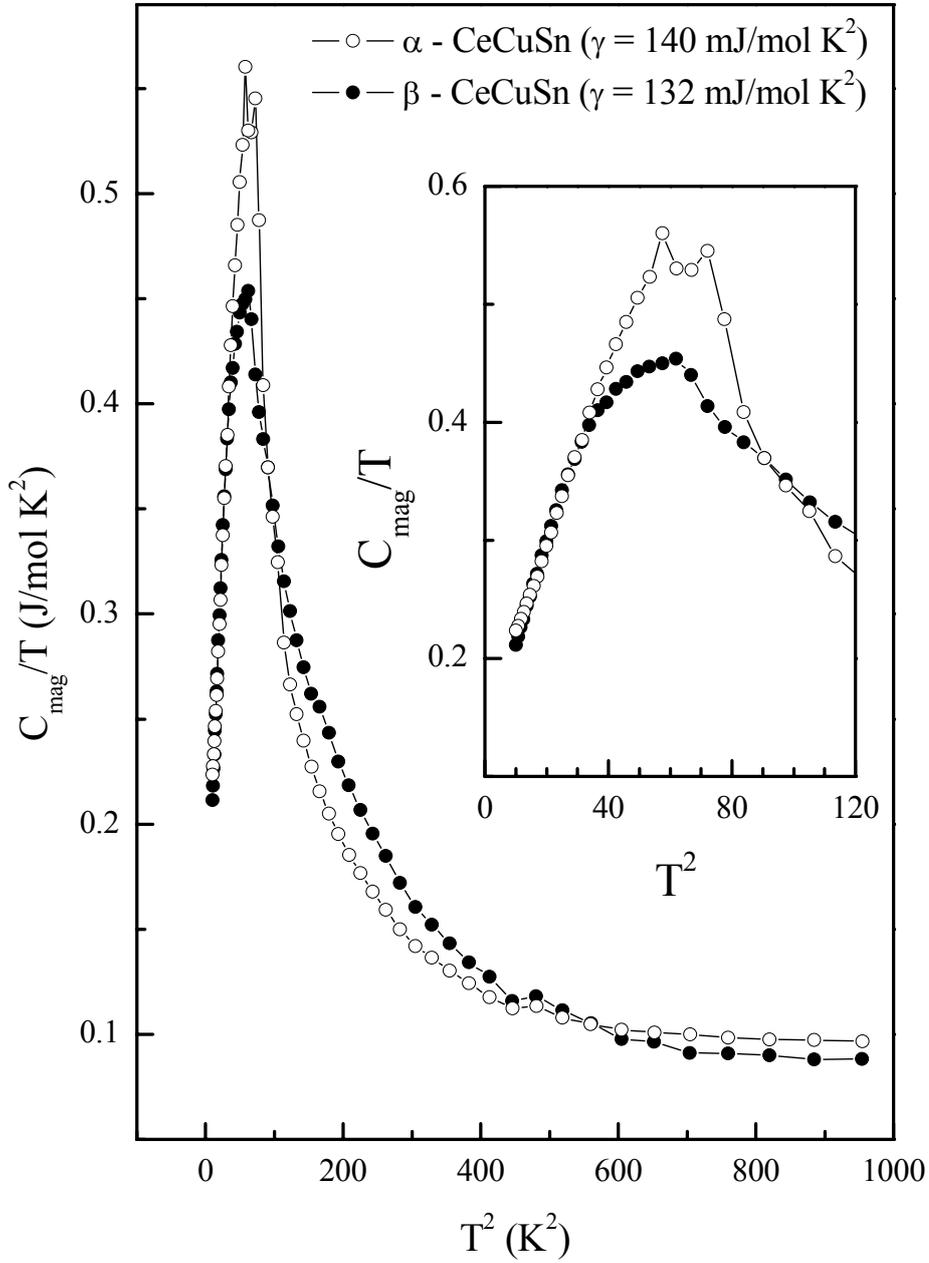

Fig. 8. $C_{mag}$ of α- and β-CeCuSn plotted as $C_{mag}/T$ vs. $T^2$ in the full scale of measurement, to show the anomalies observed at different temperatures. $C_{mag}/T$ vs. $T^2$ is shown in an expanded form in the inset to highlight the $T^3$ behaviour of $C_{mag}$ in the ordered region.

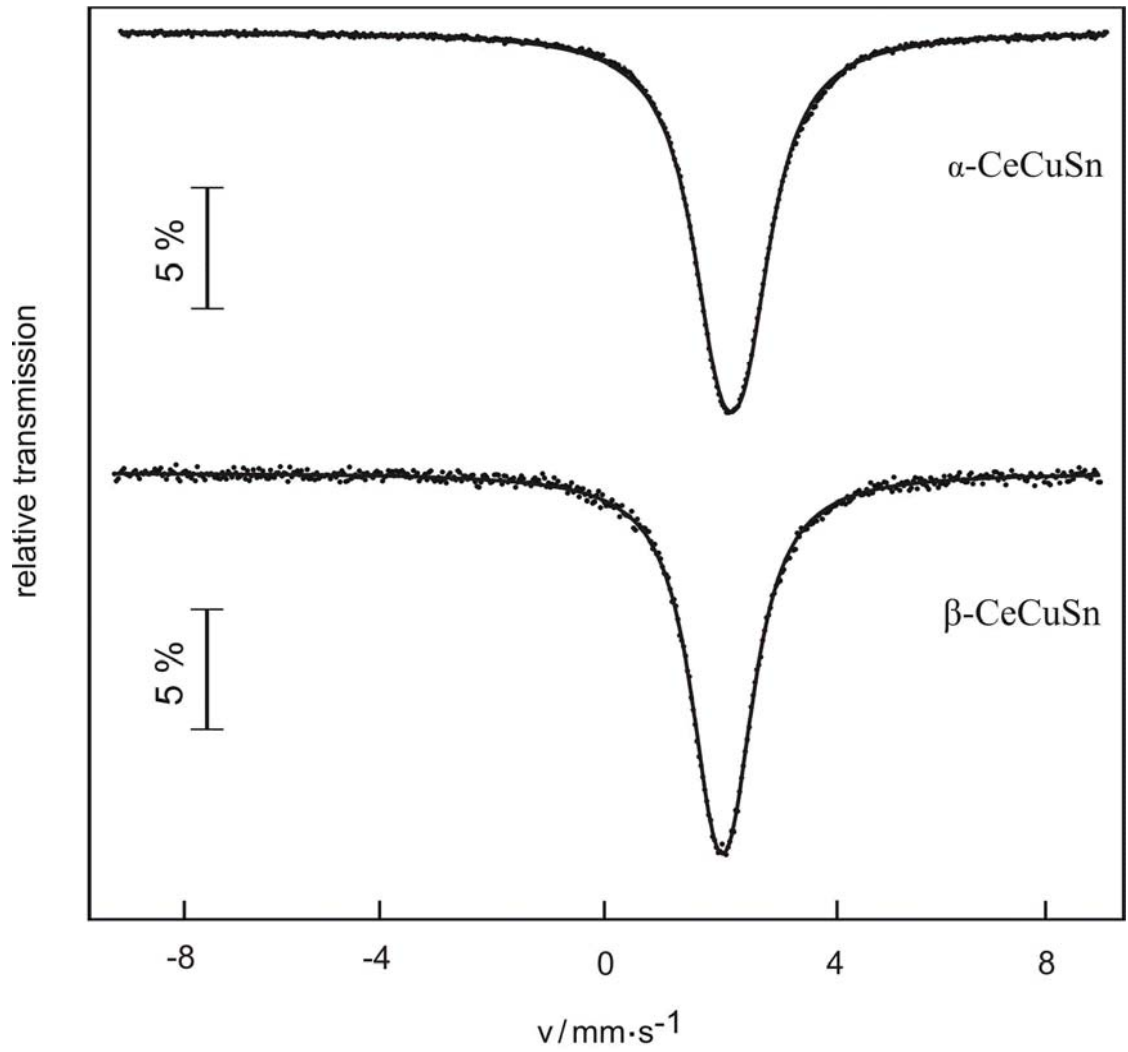

Fig. 9 Experimental (data points) and simulated (continuous line) $^{119}$Sn Mössbauer spectra of α- and β-CeCuSn at 78 K.

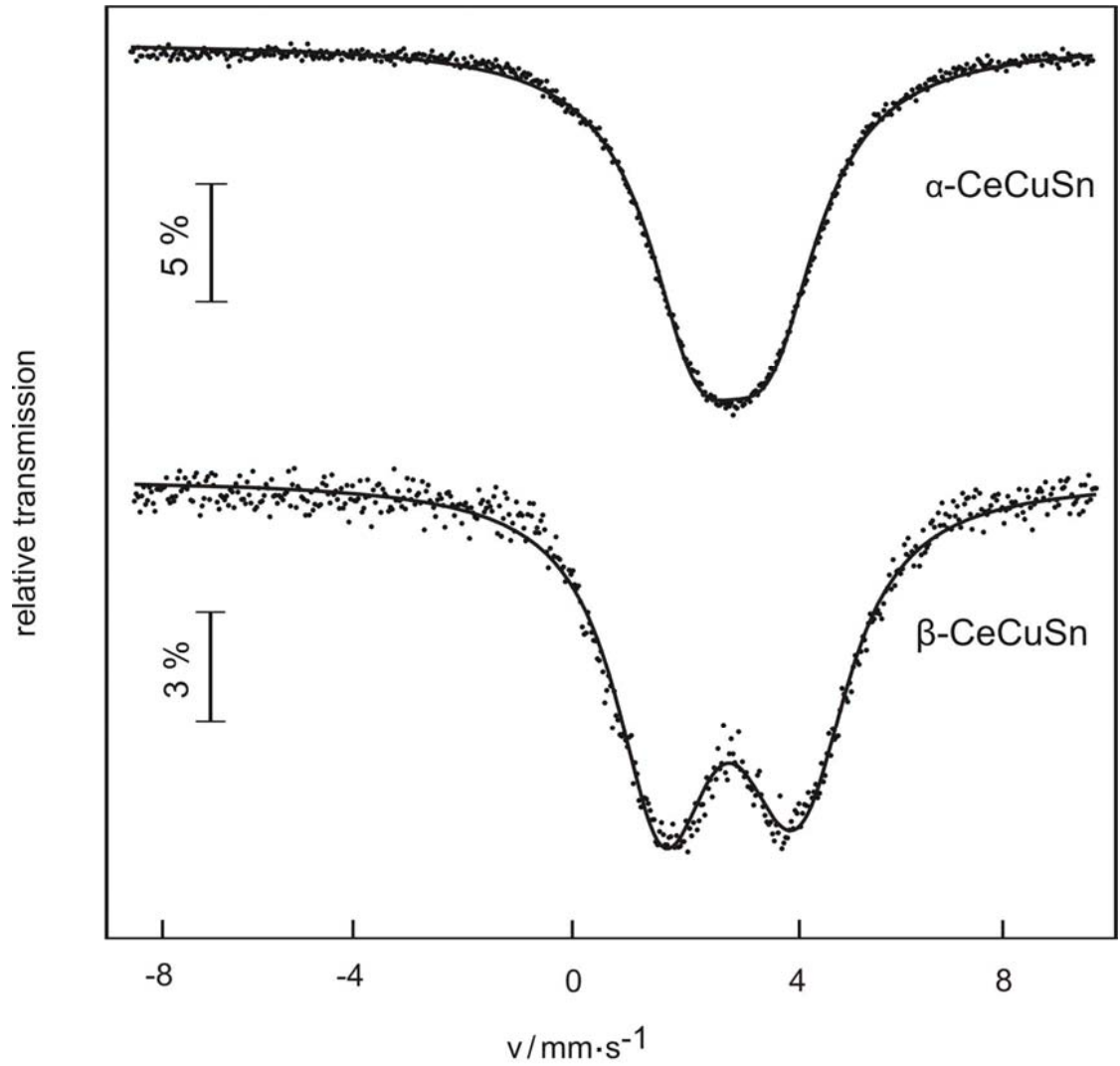

Fig. 10 Experimental (data points) and simulated (continuous line) $^{119}$Sn Mössbauer spectra of α- and β-CeCuSn at 4.2 K.